# Muon Collider Lattice Concepts

**Y. Alexahin[*], E. Gianfelice-Wendt and V. Kapin**

*Fermi National Accelerator Laboratory,*
*Batavia 60510 IL*
*E-mail*: alexahin@fnal.gov

ABSTRACT: A Muon Collider poses a number of challenging problems in the lattice design – low $\beta*$, small circumference, large physical and dynamic aperture – which must be solved in order to realize the unique opportunities it offers for the high-energy physics. This contribution presents basic solutions which make it possible to achieve the goals for both the energy frontier collider and the Higgs factory with $Nb_3Sn$ magnet parameters.

KEYWORDS: circular collider, Higgs factory, accelerator lattice, interaction region, chromaticity correction, negative momentum compaction.

---

[*] Corresponding author.

## Contents

## 1. Introduction

Muon Collider (MC) promises unique opportunities both as an energy frontier machine and as a factory for detailed study of the Higgs boson and other particles. However, in order to achieve a competitive level of luminosity a number of demanding requirements in the collider optics should be satisfied arising from the short muon lifetime and the relatively large values of the transverse emittance and momentum spread in muon beams that can realistically be achieved with ionization cooling. These requirements are aggravated by limitations on maximum operating magnetic fields as well as by the necessity to protect superconducting magnets and collider detectors from muon decay products. The design of the collider optics, magnets and machine-detector interface (MDI) is closely intertwined.

     In this paper we present the solutions for collider lattices. The magnet engineering and MDI problems will be addressed separately.

## 2. Lattice Challenges and Solutions

There is a number of requirements that are either specific to or more challenging in the case of a muon collider:

- Low beta-function at the IP: $\beta*$ values of a few millimeters are considered for muon colliders in the c.o.m. energy range 3-6 TeV.
- Small circumference $C$ to increase the number of turns (and therefore possible interactions) the muons make during their lifetime.
- High number of muons per bunch: $N_\mu \sim 2\cdot10^{12}$ and higher is envisaged.
- Protection of magnets from heat deposition and detectors from backgrounds produced by secondary particles.

Other requirements specific to either a high energy MC or the Higgs factory and be analysed in the subsequent sections. Subsection 2.1 considers lattice design based on Nb$_3$Sn magnets for collision energies $E_{\text{c.o.m.}}$= 1.5 TeV and 3 TeV and preliminary design of the $E_{\text{c.o.m.}}$= 6 TeV Interaction Region (IR) based on HTS magnets, Subsection 2.2 is devoted to the Higgs Factory lattice design.

There is an important (and beneficial) consequence of the short muon lifetime, ~ 2000 turns: high-order resonances have little chance to manifest themselves. Other sources of diffusion like IBS or residual gas scattering are also too weak to produce a halo, so if the muon beams are pre-collimated, e.g. at $3\sigma$, before injection in the collider ring their distribution is likely to stay bounded by a close value. This would relax the requirements on the dynamic aperture and on the efficiency of the halo removal from the ring. By this reason collimation sections were not included in the lattice design.

## 3. Lattices for High-Energy Colliders

In order not to lose much in luminosity due to the hour-glass effect (see e.g. ref. [1]) the bunch length should be small enough: $\sigma_z \leq \beta*$. With the longitudinal emittance expected from the final cooling channel, this will make the momentum spread $\sigma_p/p \sim 0.001$, which is an order of magnitude higher than in hadron colliders like the Tevatron and LHC. Therefore a high energy MC must have a large momentum acceptance and – to obtain small $\sigma_z$ with a reasonable RF voltage – a low momentum compaction factor $\alpha_c \sim 10^{-5}$.

There are more requirements that are specific to a high energy MC. First of all, the beam-beam effect is expected to be close to the limit, due to the necessity to put all available muons in just one bunch of each sign. This imposes constraints not only with regard to particle stability but also with respect to possible violation of aperture restrictions due to the dynamic beta effect.

Also, for beam energies $E > 2$ TeV there is a peculiar requirement for absence of straights longer than ~0.5 m in order not to create "hot spots" of neutrino radiation [2]. As a consequence quadrupoles must have a dipole component to spread out the decay neutrinos. This creates difficulties in the $\beta*$-tuning sections which must allow for $\beta*$ variation in a wide range, while maintaining dispersion closure.

In the following subsections we consider the lattice design based on Nb$_3$Sn magnets for collision energies 1.5 TeV and 3 TeV. The design parameters are given in the summary table of Section 5.

### 3.1 Interaction region

The requirements of large momentum acceptance and small $\beta*$ make correction of the Interaction Region (IR) chromaticity a challenging problem. It can be solved by using the concept of "local" correction with a few sextupoles placed not far from the IR quadrupoles. For them to work a significant dispersion has to be generated by dipoles and /or dipole components in IR quadrupoles.

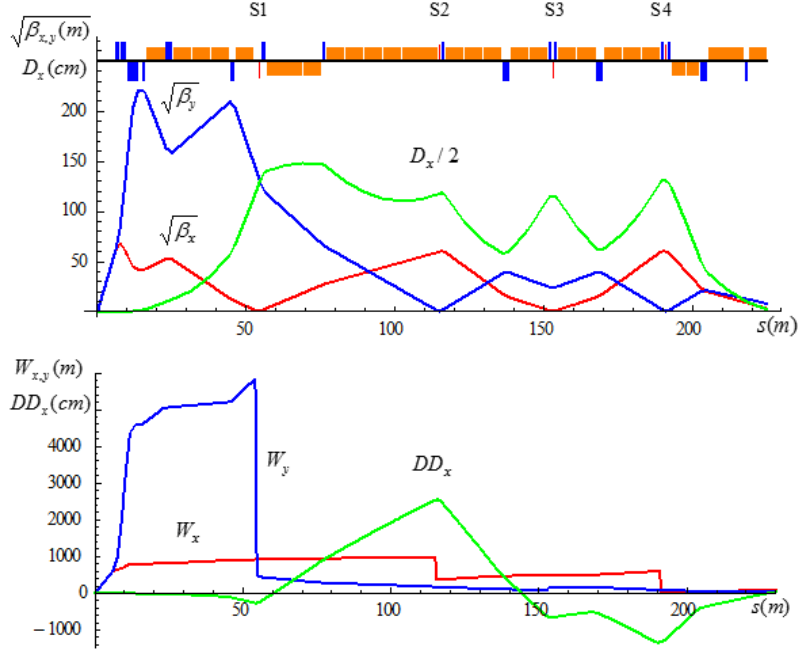

Figure 1: Layout and optics functions in the IR and CCS (top) and chromatic functions in MAD notations over the same region (bottom) of $E_{c.o.m.}$= 1.5 TeV MC with $\beta^*$=1 cm.

Another important role of these dipoles is in sweeping away from the detector the charged secondary particles created by decay in incoming muon beam. There were designs [3], [4] with dipoles placed close to the IP ("dipole first" option) but their usefulness was not obvious. Here we consider designs with IR quadrupoles that have a modest (~2 T) dipole component, with the exception of those nearest to IP: secondary particles deflected there would have a high probability of hitting the detector rather than the absorber. The IR quadrupole dipole component contribution to the dispersion is not significant, the latter is mostly generated by strong dipoles following the IR multiplets.

Figure 1 presents the layout and the optics functions in the IR and Chromaticity Correction Section (CCS) of a $E_{c.o.m.}$= 1.5 TeV Muon Collider [5] for $\beta^*$=1 cm. With a doublet Final Focus (FF) it was possible to achieve the design goal parameters with moderate strength $Nb_3Sn$ quadrupoles.

For higher energies and/or smaller $\beta^*$ the doublet FF would require the use of HTS magnets as in the $E_{c.o.m.}$= 6 TeV design with $\beta^*$=1 cm [6]. However, to keep luminosity increasing with energy as $E^2$ a smaller $\beta^*$ value is necessary.

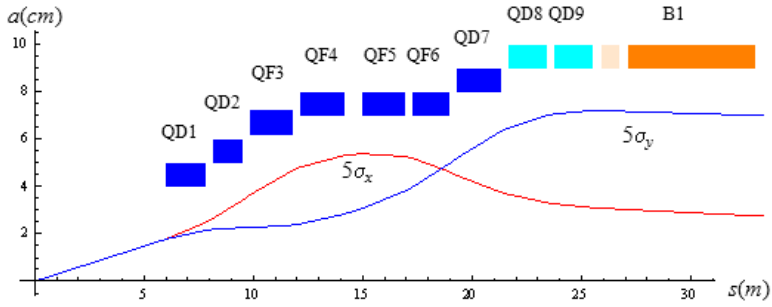

Figure 2: Triplet FF quadrupole apertures and $5\sigma$ beam envelopes for $E_{c.o.m.}$= 3 TeV and $\beta^*$ = 5mm. Defocusing magnets with 2 T dipole component are shown in cyan.

Table 1: $E_{c.o.m.}$= 3 TeV IR Magnet Parameters.

| Parameter | QD1 | QD2 | QF3 | QF4-6 | QD7 | QD8-9 | B1 |
|---|---|---|---|---|---|---|---|
| Aperture (mm) | 80 | 100 | 124 | 140 | 160 | 180 | 180 |
| Gradient (T/m) | -250 | -200 | 161 | 144 | 125 | -90 | 0 |
| $B_{dipole}$ (T) | 0 | 0 | 0 | 0 | 0 | 2 | 8 |
| Length (m) | 1.85 | 1.40 | 2.00 | 1.70 | 2.00 | 1.75 | 5.80 |

An obvious solution is to use a triplet FF. Figure 2 shows the IR design for $E_{c.o.m.}$= 3 TeV and $\beta^*$=0.5 cm [7], which is based on Nb$_3$Sn quadrupoles with the parameters listed in Table 1. The quadrupole inner radii satisfy the requirement: $R > 5\sigma_{max} + 2$ cm, which guarantees that the beam will be in a good field region and provides enough space for absorbers. The gradients were chosen so that the field at the inner bore radius does not exceed 10T.

The order of focusing and defocusing quadrupoles was chosen to minimize the horizontal beta-function at the B1 dipole used for dispersion generation. In the result the quadrupoles that are closest to the IP and farthest from it are defocusing while the inner group is focusing.

There is a certain disadvantage with the inner group of quadrupoles being focusing: the dipole component in strong focusing quadrupoles is not as efficient in sweeping away secondary particles as in defocusing quadrupoles, so it was not introduced in the inner group, despite its favorable location for that purpose.

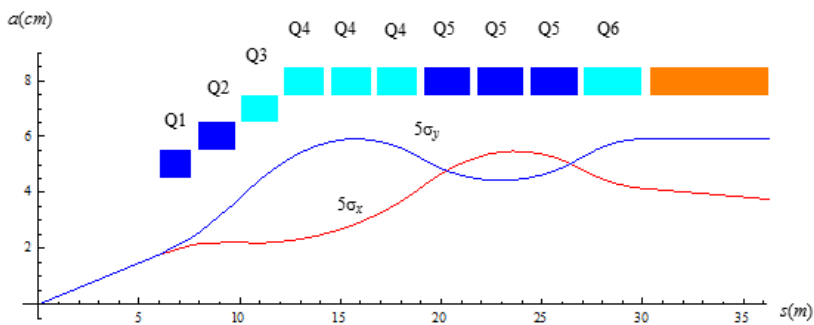

Figure 3: Quadruplet FF quadrupole apertures and $5\sigma$ beam envelopes for $E_{\text{c.o.m.}}$ = 3 TeV and $\beta^*$ = 5mm. Defocusing magnets with 2 T dipole component are shown in cyan.

The contradiction between the requirements of both the second and the last from IP groups of quadrupoles being defocusing is naturally resolved in a quadruplet FF. Such a design – with 12 T limit on the total field at the inner bore of the magnets – is presented in Fig. 3. The final choice between the triplet and quadruplet versions can be made only after careful study of energy deposition and detector backgrounds.

### 3.2 Chromaticity correction

It is necessary to correct the IR quadrupole chromaticity in such a way that the dynamic aperture remain sufficiently large, and does not suffer much from strong beam-beam effects and magnet errors.

To achieve these goals, a solution was proposed in the past based on special Chromatic Correction Sections (CCS) with compensated spherical aberrations [8]. Each CCS includes two sextupoles separated by a –I transformation so that their nonlinear kicks cancel out. There is an independent CCS for each transverse plane making the total of four chromaticity correction sextupoles on each side of the IP. This approach has led to a number of muon collider designs; the best performance among early versions was demonstrated by a 4 TeV c.o.m. collider design by K. Oide [9].

There are two major problems with CCS. First, they are sources of significant chromaticity themselves so that the required integral strength of the two sextupoles in a CCS is higher than with a single sextupole correction. In the result the higher order effects in the sextupole strength are greatly enhanced, limiting the dynamic aperture (especially vertical) by comparable or even lower values.

The authors of the 6 TeV design [6] have overcome this difficulty by adding weak compensating sextupoles at some (small) distance from the main sextupoles in the vertical CCS thus recovering the vertical dynamic aperture.

The second problem is the sensitivity to magnet field errors and misalignments which increases with the increased number and strength of elements at high beta locations. To reduce such sensitivity a three-sextupole scheme was proposed in ref. [5], where the vertical chromaticity is corrected with a single sextupole (S1 in Figure 1) placed at the same vertical phase advance as the FF quadrupoles. The spherical aberrations are reduced with the help of a low horizontal beta-function at its location. For horizontal chromaticity correction a CCS is still necessary since smallness of $\beta_y$ at a normal

– 5 –

sextupole location is beneficial but does not suppress horizontal aberrations. In Figure 1 the horizontal CCS spans the region from S2 to S4.

Sextupoles S2, S3 and S4 form a chromatic three-bump to compensate the second order dispersion $DD_x = dD_x/d\delta = d^2x/d\delta^2$ ($\delta = \Delta E / p_0 c$) generated by the upstream sources (bottom plot of Fig. 1). Since the strength of S2 and S4 should be equal for cancellation of aberrations the net effect on $DD_x$ is achieved by creating a difference in dispersion $D_x$ at their locations. The locations of S1-S4 are convenient for placing correctors of higher order dispersion, chromaticity of tunes and beta-functions. In particular, octupoles at S2 and S4 locations are used for correction of the third order dispersion at IP and the second order chromaticity.

### 3.3 Flexible momentum compaction arc cell

The interaction region produces a large positive contribution to the momentum compaction factor $\alpha_p$ which must be compensated by a negative contribution from the arcs.

In ref. [10] a new version of the so-called Flexible Momentum Compaction (FMC) arc cell was proposed which allowed for independent control of all important parameters: tunes, chromaticities, momentum compaction factor and its derivative with momentum.

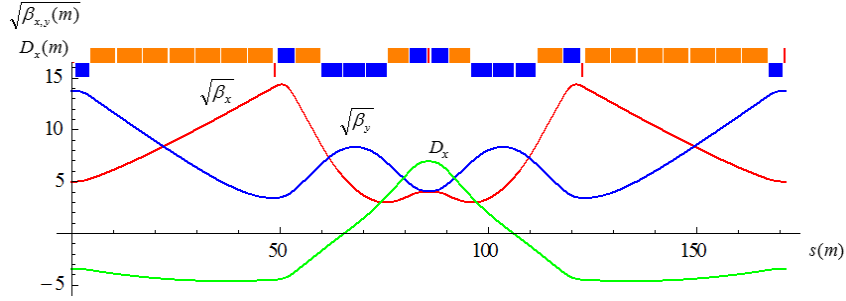

Figure 4: Layout and optics functions in $E_{c.o.m.}$= 3 TeV MC arc cell.

That design was based on separate-function magnets with rather long quadrupoles which are not suitable for high energy machine because of neutrino-induced radiation [2]. Also, simulations of energy deposition by decay electrons in magnets [11] showed that the large vertical displacement these electrons can obtain in quadrupoles makes the choice of an open-midplane dipole design ineffective.

Both of the above-mentioned problems can be alleviated by employing combined-function magnets [7]. Magnet parameters for the design presented in Figure 4 are as follows: focusing magnets $B = 8$ T, $G \leq 87$ T/m, $L = 4$ m, defocusing magnets $B = 9$T, $G = -35$ T/m, $L \leq 5$ m, and pure dipoles $B = 10$ T, $L \leq 6$ m. The momentum compaction factor for a stand-alone cell is $\alpha_c = -0.004$. Each arc consists of six such cells and two dispersion suppressors. The betatron phase advance is 300° in both planes to ensure cancellation of spherical aberrations. Though the sextupoles for chromaticity correction are interleaved they are too weak to noticeably affect the dynamic aperture.

At the same locations multipoles for higher order chromaticity correction can be placed. Unlike the multipoles at S1-S4 locations, they (practically) do not affect the off-momentum beta-beating and can be used in addition to them for correction of the chromaticity of the tunes.

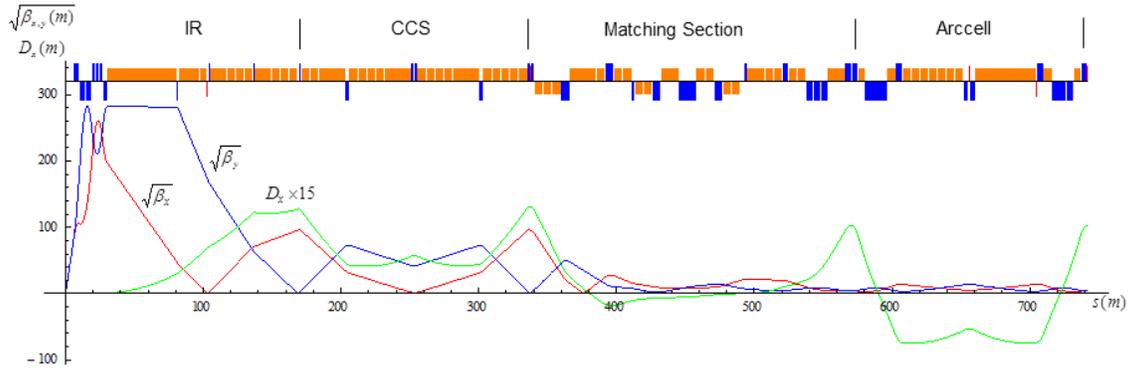

Figure 5: Layout and optics functions from IP to the end of the first arc cell in 3TeV MC with quadruplet FF ($\beta^*$=5mm).

### 3.4 Beta-tuning section

The design should be flexible enough to allow a wide range of $\beta^*$ values, for a number of reasons. First, it is easier to start machine running with large $\beta^*$. Second, there is an uncertainty in the muon beam emittance that can be obtained in the ionization cooling channel. With higher emittance $\beta$-functions in the IR magnets should become smaller in order to accommodate the beam inside the available aperture and in the result $\beta^*$ has to be larger. Conversely, with lower emittance smaller values of $\beta^*$ are allowed.

To maintain chromaticity correction, the IR and CCS parameters should not change. Therefore the $\beta^*$ variation should be produced by quadrupoles in the adjacent IR-to-Arc matching section. There will be four such sections in the ring, which will also serve as utility sections. They should satisfy a number of requirements:
- allow $\beta^*$ variation in a wide range (e.g. 3 mm - 3 cm),
- have no long straights without bending field, and
- provide spaces with low $\beta$-functions and dispersion for RF cavities.

The first two requirements are difficult to reconcile: $\beta_x$ variation at a bend will change dispersion; trying to adjust the bending field will change the orbit. A possible solution to this problem is the use of a chicane with adjustable bending field which does not perturb the orbit outside and changes the total orbit length only slightly [7].

Figure 5 shows the layout and optics functions in $E_{c.o.m.}$= 3 TeV MC from IP to the end of the first arccell with the $\beta^*$-tuning section stretching from 340 m to 500 m. The rest of the matching section is occupied by the arc dispersion suppressor.

## 3.5 Lattice performance

Performance of the 1.5 TeV MC lattice was discussed in ref. [11]. Later work [12] considered the effect of fringe fields and systematic magnet errors. The fringe fields were shown to reduce the vertical dynamic aperture by almost a factor of two but no attempt was made to compensate them. Previous experience [9] showed that this can be done with the help of multipole correctors. The main result of ref. [12] was the demonstration of a strong adverse effect of the systematic field errors – especially of the decapole component – in the open-midplane dipoles. Such dipoles were designed (see ref. [11]) in the hope to drastically reduce heat deposition from muon decay products.[1] There were no studies yet of the effect of fringe fields and magnet errors for the 3 TeV MC.

In the 1.5 TeV MC design the tunes were chosen just above half-integer to reduce closed orbit sensitivity to misalignments and dipole errors. For the 3 TeV MC two working points were considered with the tunes above either half-integer or integer values. Figure 6 shows the dependence of the tunes, $\beta$-functions at the IP and the momentum compaction factor on the relative momentum deviation $\delta_p$ in the ideal 3 TeV MC lattice with the near-integer working point. The stable momentum range exceeds ±6$\sigma$. It is limited by uncorrected fourth order chromaticity. The very low central value of the momentum compaction factor, $\alpha_c(0) = -2.15 \cdot 10^{-6}$, can be increased if necessary for operational considerations.

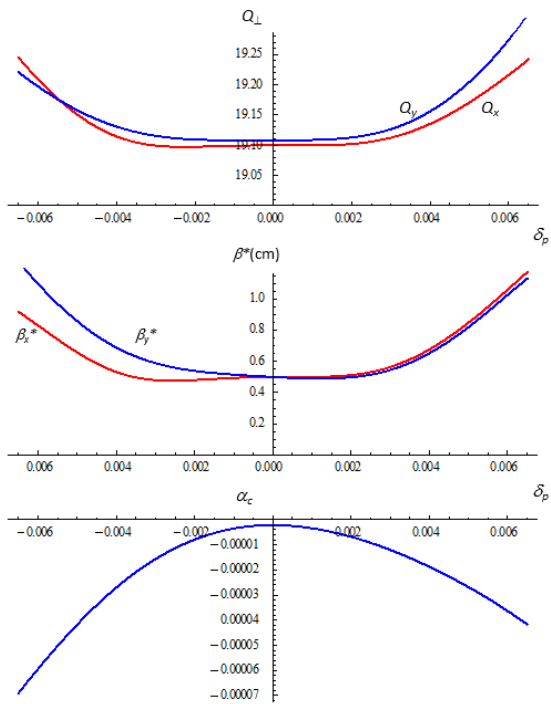

Figure 6. 3TeV MC bare lattice parameters vs. relative momentum deviation $\delta_p$ ($\beta^* = 5$mm).

Figure 7 shows the 3 TeV MC 2048 turns dynamic aperture in the plane of the initial particle coordinates at the IP $x_{in}$, $y_{in}$ for indicated values of constant $\delta_p$ calculated with beam-beam interaction off (solid lines) and on (dashed line) using the MADX PTC_TRACK routine and the MAD8 TRACK LIE4 option, respectively.

---

[1] This hope was refuted by MARS simulations [13] which showed that decay electrons are deflected onto the cold mass by fields in the gap.

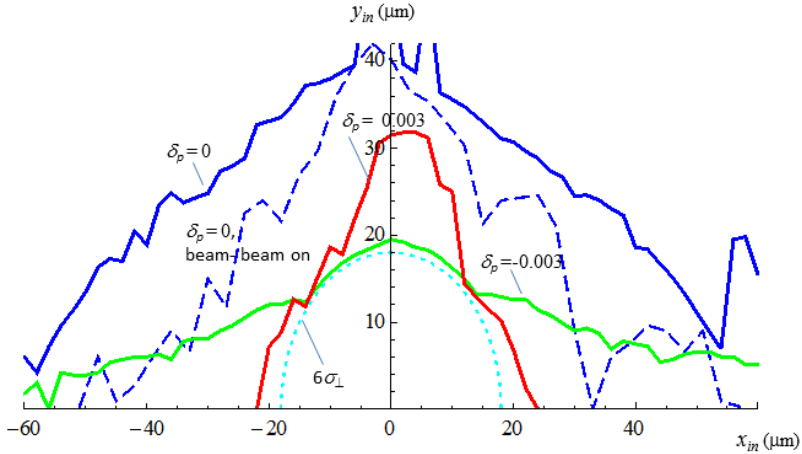

Figure 7: The 3 TeV MC dynamic aperture in the plane of initial particle coordinates for $\beta^*$ = 5 mm. Solid and dashed blue lines: $\delta_p$ = 0 without and with beam-beam interaction. Green and red lines: $\delta_p$ = -0.003 and $\delta_p$ = 0.003 respectively without beam-beam interaction. Dotted line shows the $6\sigma$ beam ellipse.

Since the beam size at the IP is $\sigma_\perp$ = 3 μm, the off-momentum dynamic aperture exceeds $6\sigma$ in absence of beam-beam interaction. It should be noted that with this working point the phase advance between two S1 sextupoles in the same superperiod – e.g. one downstream of the first IP and the other upstream of the second IP – is close to an odd multiple of $\pi$ in both planes leading to cancellation of the horizontal kicks exerted by these sextupoles on particles with large vertical displacement. This explains the observed large dynamic aperture.

In the case of half-integer tunes, there is no such cancelation, but the on-momentum dynamic aperture exceeds $6\sigma$ with the beam-beam interaction both off and on. The off-momentum dynamic aperture is smaller, especially with the beam-beam interaction on.

The above results show that the three-sextupole chromaticity correction scheme can provide sufficient dynamic aperture and momentum acceptance of the ideal lattice. The remaining problem is correction of the effects of fringe fields and magnet errors – both random and systematic – which should be addressed when the work on the muon collider is resumed.

### 3.6 Beam-beam effect

For high energy muon colliders a very large beam-beam tuneshift of $\approx$ 0.1/IP is envisaged. The beam-beam interaction not only introduces strong nonlinearities but also changes the "cross-talk" of the lattice nonlinear elements across the IP making it necessary to take the beam-beam effect into account at this stage of lattice design.

The dashed curve in Figure 7 demonstrates the effect of the beam-beam interaction in the weak-strong approximation on a particle with $\delta_p$ = 0 in the center of the bunch. In this approximation the "dynamic beta effect" reduces $\beta^*$ from 5 to 3 mm, so in units of $\sigma_\perp$ the dynamic aperture becomes even larger. With $\delta_p \neq 0$ the dynamic aperture decreases becoming only marginally sufficient at $\delta_p$ = -0.0025. Some adjustments of the tunes and nonlinear corrector settings are necessary.

The dynamic beta effect in the strong-strong approximation is even more pronounced. Figure 8 shows the results of calculations for the integer tunes case in the linear approximation of the beam-beam force using A. Netepenko's Mathematica code [10]. In these calculations a Gaussian longitudinal profile was assumed and represented by a number of slices (23 per bunch) according to the Zholents-Shatilov algorithm [14].

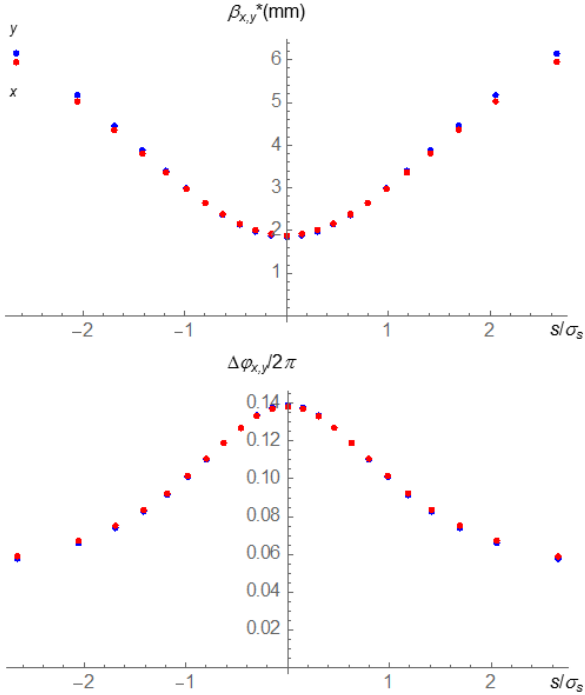

Figure 8: Dynamic $\beta^*$ (top) and the beam-beam tuneshift per IP (bottom) seen by particles along the bunch in the case of tunes just above integer.

As Figure 8 shows, in the strong-strong case $\beta^*$ for the central slice is reduced from 5mm to less than 2mm while the tuneshift increases from 0.09 to 0.14 per IP. Away from the center $\beta^*$ goes up while the tuneshift decreases – just opposite to the case of half-integer tunes [15]. The dynamic beta effect increases $\beta$-functions in the Final Focus multiplet but not as much as would be required to achieve the same $\beta^*$ by retuning the bare lattice.

In the following we will assume that $\beta^*$ in the bare lattice is increased by the amount necessary to obtain the design value in the presence of the dynamic beta effect. This will certainly alleviate the above-mentioned problem with the off-momentum dynamic aperture.

### 3.7 Preliminary IR design of the 6 TeV MC

The 1.5 TeV and 3 TeV muon collider designs assumed the presently available $Nb_3Sn$ magnet technology. In the future the HTS technology can mature enough to become practical for large-scale applications, making it possible to achieve 16 T pole tip fields in quadrupoles and 20 T fields in dipoles.

Figure 9 shows the 6 TeV MC interaction region design based on such assumption. The design goals were: $\beta^* = 3$ mm, 10 m distance from IP to the first quad, at least 3 T dipole component in quadrupoles (with exception for the first one) to sweep away charged secondary particles. The magnet inner bore radius was constrained by the requirement $IR > 5\sigma_\perp^{(max)} + 3$ cm which also guarantees good field quality in the beam region. Magnets were cut in pieces shorter than 6 m to insert protecting masks.

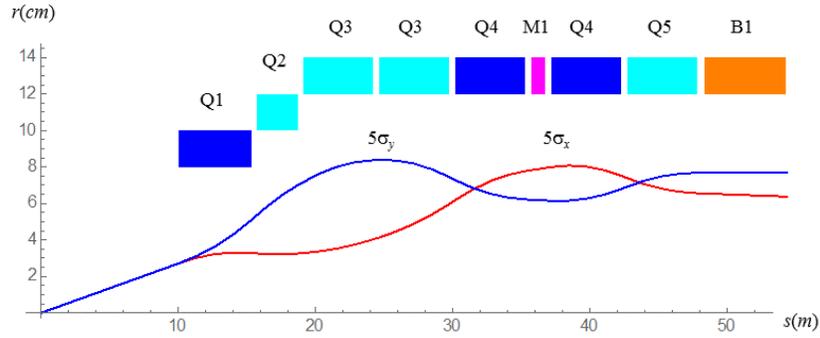

Figure 9: 6 TeV IR quadrupoles aperture and $5\sigma$ beam envelopes for $\beta^* = 3$ mm. Defocusing magnets with up to 5 T dipole component are shown in cyan.

The requirements on the IR magnets are summarized in Table 2. Despite high field and large aperture these quadrupoles look less challenging than magnets for the Higgs Factory (see next Section) based on the $Nb_3Sn$ technology.

Table 2. Parameters of the 6 TeV Final Focus quadrupoles

| *Parameter* | *Q1* | *Q2* | *Q3* | *Q4* | *Q5* |
|---|---|---|---|---|---|
| ID (mm) | 160 | 200 | 240 | 240 | 240 |
| G (T/m) | 200 | -125 | -100 | 103 | -78 |
| $B_{dipole}$ (T) | 0 | 3.5 | 4.0 | 3.0 | 6.0 |
| L (m) | 5.3 | 3.0 | 5.1 | 5.1 | 5.1 |

## 4. Specifics of the Higgs Factory Lattice

There are a number of advantages of a muon collider Higgs factory [16], and among them a high cross-section for Higgs boson production in the *s*-channel and the possibility of obtaining a sufficiently low muon beam energy spread to directly measure the Higgs boson peak width, which is expected to be ~4MeV.

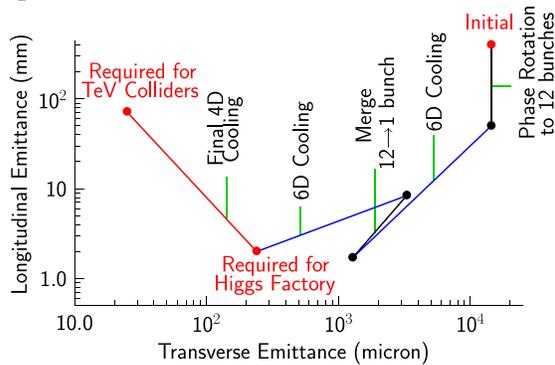

Figure 10: Ionization cooling of muons required for the Higgs factory and high-energy muon colliders.

To obtain a low energy spread, D. Neuffer proposed stopping muon cooling where the longitudinal emittance reaches its minimum (see Fig. 10) leaving the transverse emittance relatively high [16]. As a consequence, quite small values of the beta-function (a few cm) at the Interaction Point are required to achieve a sufficiently high luminosity. This results in a large beam size in the Final Focus quadrupoles. This factor, as well as the requirements for magnet and detector protection from muon decay products, requires very large aperture for the IR magnets posing challenging engineering constraints and presenting beam dynamics issues from magnet fringe fields and the body field errors.

The overall lattice design for the Higgs factory strongly differs from that of a high energy muon collider. In the latter case a small as possible momentum compaction factor is needed to obtain small bunch length with large longitudinal emittance, while in the Higgs factory the main concern is preservation of a small beam energy spread. There are a number of effects which can increase energy spread, among them: microwave instability, longitudinal beam-beam effect, path-lengthening due to transverse oscillations. Their impact is discussed in the following subsections.

### 4.1 Microwave instability

For a rough estimate we can use the Keil-Schnell criterion (see e.g. Ref. [1] p.118)

$$\left|\frac{Z_\parallel}{n}\right| \leq \frac{2\pi E |\alpha_c|}{eI_{peak}} \left(\frac{\sigma_E}{E}\right)^2 \qquad (1)$$

Table 3: Higgs Factory reference parameters

| Parameter | Unit | Value |
|---|---|---|
| Beam energy | GeV | 63 |
| Transverse emittance, $\varepsilon_{\perp N}$ | ($\pi$)mm·rad | 0.3 |
| Longitudinal emittance, $\varepsilon_{\parallel N}$ | ($\pi$)mm·rad | 1.0 |
| Number of bunches/beam | - | 1 |
| Number of muons/bunch | - | $2\times10^{12}$ |
| Number of IPs | - | 1 |
| Circumference | m | 300 |
| Beam energy spread | % | 0.003 |
| Bunch length | cm | 5.6 |
| $\beta^*$ | cm | 2.5 |
| Momentum compaction factor $\alpha_c$ | - | 0.079 |
| Beam-beam parameter | - | 0.007 |

With the muon beam parameters from Table 3, $I_{peak} = 0.68$ kA and for a typical value of $|Z_\parallel/n| \sim 0.1\ \Omega$ we obtain a very high threshold value $|\alpha_c| > 0.19$. Criterion (1) is over-restrictive for bunched beams, but a similar estimate for $\alpha_c$ can be obtained just from the requirement that the RF voltage was significantly higher than the bunch self-fields without increasing the energy spread.

### 4.2 Longitudinal beam-beam effect

Another effect which can play a role in the Higgs factory case is the longitudinal beam-beam effect first described in [17].

When a witness particle interacts with a thin slice of opposing bunch its energy changes depending on longitudinal position $s_{coll}$ of the actual collision point

$$\Delta E = \frac{e^2 N_s}{2\beta_\perp} \frac{d\beta_\perp}{ds}\bigg|_{s=s_{coll}} \qquad (2)$$

where $N_s$ is the number of particles in the slice. Integrating over the strong bunch length we obtain dependence of the energy change per pass on particle longitudinal position shown in Fig. 11. This energy change is comparable to that from RF (see Table 3) and is strongly nonlinear so that a higher harmonic RF may be needed for correction.

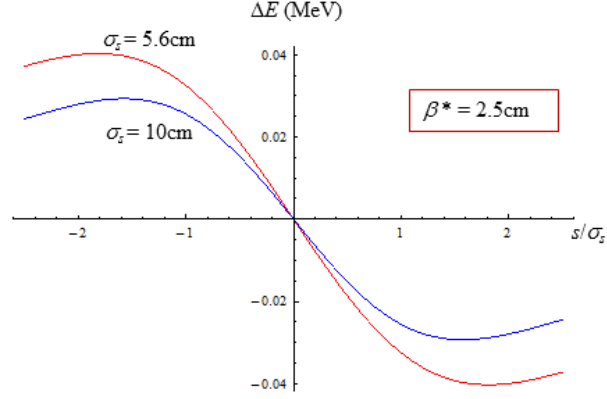

Figure 11: Longitudinal kick vs. particle position in the bunch for the reference parameters of Table 3.

### 4.3 Path-lengthening effect

It may seem that, due to a small energy spread, chromaticity correction is not needed in the Higgs factory. However, there other effects which may require chromaticity correction, in particular the path length dependence on betatron amplitude [18], [19] which translates into additional energy spread:

$$\frac{\Delta E}{E} \approx \frac{1}{\alpha_c R}(Q'_x I_x + Q'_y I_y) \qquad (3)$$

where $Q_x'$, $Q_y'$ are chromaticities (tune derivatives with respect to relative momentum deviation), $I_x$, $I_y$ are action variables and $R$ is average machine radius. With uncorrected chromaticity in both planes, $Q_\perp' \sim -100$ (actually it can be larger in magnitude than that) and with $\alpha_c = 0.1$ we would have the average energy shift

$$\left\langle \frac{\Delta E}{E} \right\rangle = \frac{2 Q'_\perp \varepsilon_{\perp N}}{\alpha_c R \gamma} \approx -2 \cdot 10^{-5} \qquad (4)$$

while contribution to the energy spread would be of the same order as the goal value. This shows the necessity to correct chromaticity. Chromaticity correction is also needed from operational point of view.

### 4.4 Higgs factory lattice design

For the IR design we use the same concepts as for the 3 TeV collider design: quadruplet Final Focus (FF) and the 3-sextupole chromaticity correction scheme. Parameters used in the lattice design are given in Table 3. We will refer to them as the reference parameters. Based on them the parameter sets for the initial operation and an upgrade were determined.

The IR design assumes a 3.5 m distance from Q1 to the IP and a quadrupole bore diameter of ($10\sigma_{max} + 30$) mm. Figure 12 shows the $5\sigma$ beam envelopes and the required magnet inner radii. Table 4 gives the IR magnet parameters. Splitting Q2 in two parts allows for a mask in between.

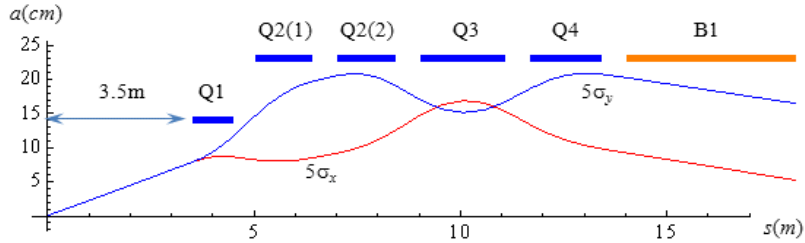

Figure 12: Higgs factory IR quadrupoles aperture and $5\sigma$ beam envelopes for $\beta^* = 2.5$ cm.

Table 4: Higgs Factory IR Magnet Specifications

| *Parameter* | *Q1* | *Q2* | *Q3* | *Q4* | *B1* |
|---|---|---|---|---|---|
| Aperture (mm) | 270 | 450 | 450 | 450 | 450 |
| Gradient (T/m) | 74 | -36 | 44 | -25 | 0 |
| Dipole field (T) | 0 | 2 | 0 | 2 | 8 |
| Magnetic length (m) | 1.00 | 1.40 | 2.05 | 1.70 | 4.10 |

The optics functions in a half ring (starting from the IP) are shown in Figure 13 for $\beta^* = 2.5$ cm. Note that with this IR design, $\beta^*$ can be varied from 1.5 to 10 cm by changing the gradients in matching sections without perturbing the dispersion function. The momentum acceptance of the ring exceeds $\pm 0.5\%$.

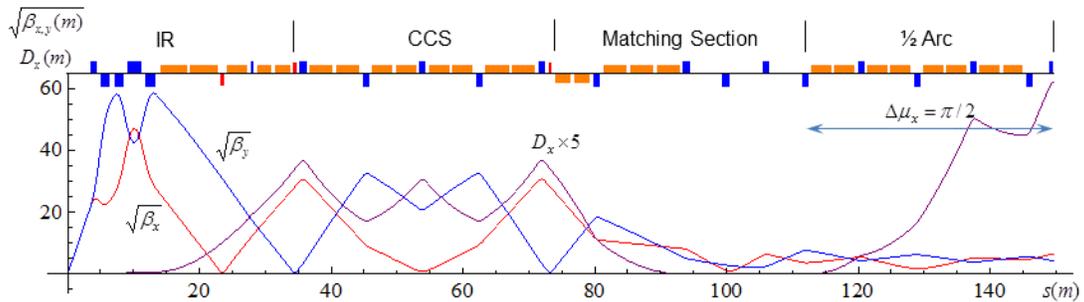

Figure 13: Layout and optics functions in half ring of the Higgs factory.

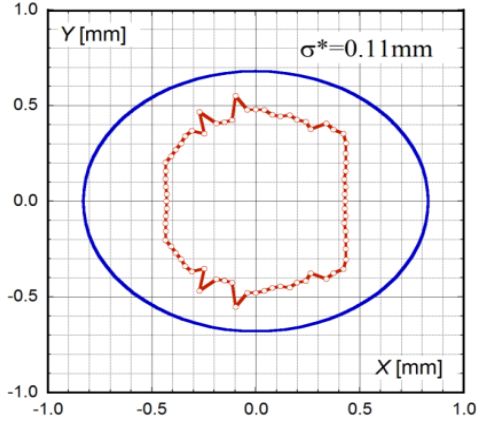

Figure 14: Higgs factory dynamic aperture in the presence of IR magnets multipole errors and projection of the magnets physical aperture (solid ellipse).

Without magnet errors the Dynamic Aperture (DA) is 8σ for $β^* = 2.5$ cm and 5σ for $β^* = 1.5$ cm. In both cases the DA significantly exceeds the physical aperture (5σ and 4σ respectively). However, the systematic field errors – especially those in IR magnets - produce a strong impact on particle dynamics. With the geometrical field harmonics calculated for the IR magnet design presented in Ref. [20], the DA is reduced by a factor of 2, to the area shown in Figure 14 which roughly corresponds to the magnet good field region.

In the above analysis the fringe-fields were not taken into account which will further reduce the DA making it necessary the use of nonlinear correctors.

## 5. Lineup of the Muon Collider Designs

Table 5 summarizes the parameters of muon collider designs for different energies considered in this paper. For the Higgs factory we included two sets of parameters: a) for the initial operation and b) the baseline set. The latter envisages twice higher number of muons per bunch and correspondingly higher longitudinal emittance, bunch length and energy spread as needed for beam stability.

Table 5: Muon collider design parameters

| *Parameter* | *Higgs Factory* | | *High Energy Muon Colliders* | | |
|---|---|---|---|---|---|
| Collision energy, TeV | 0.126 [a)] | 0.126 [b)] | 1.5 | 3.0 | 6.0* |
| Repetition rate, Hz | 30 | 15 | 15 | 12 | 6 |
| Average luminosity / IP, $10^{34}$/cm²/s | 0.0017 | 0.008 | 1.25 | 4.6 | 11 |
| Number of IPs | 1 | 1 | 2 | 2 | 2 |
| Circumference, km | 0.3 | 0.3 | 2.5 | 4.34 | 6 |
| β*, cm | 3.3 | 1.7 | 1 | 0.5 | 0.3 |
| Momentum compaction factor $α_c$ | 0.079 | 0.079 | $-1.3·10^{-5}$ | $-0.5·10^{-5}$ | $-0.3·10^{-5}$ |
| Normalized emittance, π·mm·mrad | 400 | 200 | 25 | 25 | 25 |
| Momentum spread, % | 0.003 | 0.004 | 0.1 | 0.1 | 0.083 |
| Bunch length, cm | 5.6 | 6.3 | 1 | 0.5 | 0.3 |
| Number of muons / bunch, $10^{12}$ | 2 | 4 | 2 | 2 | 2 |
| Number of bunches / beam | 1 | 1 | 1 | 1 | 1 |
| Beam-beam parameter / IP | 0.005 | 0.02 | 0.09 | 0.09 | 0.09 |
| RF frequency, GHz | 0.2 | 0.2 | 1.3 | 1.3 | 1.3 |
| RF voltage, MV | 0.1 | 0.1 | 12 | 50 | 150 |

*) The 6 TeV ring design is not completed yet, the numbers are a projection.

**Acknowledgments**

The authors are indebted to many persons for encouragement, advice and help, in particular to D. Neuffer, K. Oide, K. Ohmi, M. Palmer, R. Palmer and V. Shiltsev.

**References**

[1] A. Chao, M. Tigner, *Handbook of Accelerator Physics and Engineering*, World Sci., 1999.

[2] C. Ankenbrandt et al., *PRSTAB* **2** (1999) 081001.

[3] C.J. Johnstone, N.V. Mokhov, *Shielding the muon collider interaction region*, PAC-1997, FERMILAB-CONF-97-487-APC (1997).

[4] Y. Alexahin, E. Gianfelice-Wendt, PAC09, TH6PFP051.

[5] Y. Alexahin et al., *PRSTAB* **14** (2011) 061001.

[6] M.-H. Wang et al., IPAC-2015-TUPTY081.

[7] Y. Alexahin, E. Gianfelice-Wendt, IPAC-2012-TUPPC041.

[8] J. Irwin et al., in Proc. 1991 IEEE PAC, San Francisco, 1991, p.2058

[9] K. Oide, unpublished.

[10] Y. Alexahin, E. Gianfelice-Wendt, A. Netepenko, in Proc. IPAC10, Kyoto, 2010, p.1563

[11] N.V. Mokhov et al., in Proc. PAC11, New York, 2011, p. 2295.

[12] Y. Alexahin, E. Gianfelice-Wendt, V. Kapin, IPAC-2012-TUPPC042.

[13] N.V. Mokhov, *Review of Energy Deposition Studies*, *Workshop on Muon Collider Ring Magnets*, Fermilab, May 2011; https://indico.fnal.gov/event/4404/

[14] M. Furman, A. Zholents, T. Chen, D. Shatilov, *Comparison of Beam-Beam Code Simulations*, CBP Tech. Note-59, PEP-II/AP Note 95-04, 1996.

[15] Y. Alexahin, K. Ohmi, IPAC-2012-WEPPR004.

[16] D.V. Neuffer et al., IPAC-2013-TUPFI056.

[17] Y.A. Derbenev, A.N. Skrinsky, in Proc. *3rd All-Union Meeting on Charged Part. Acc.*, Moscow, 1972, v.1, p.386.

[18] L. Emery, in Proc. HEACC'92, Hamburg, 1992, p.1172.

[19] J.S. Berg, *Amplitude dependence of time of flight and its connection to chromaticity*, *NIM* A **570** (2007) 15-21.

[20] A.V. Zlobin et al., IPAC-2013-TUPFI061.